\documentclass[12pt]{iopart}
\usepackage{graphics}
\usepackage{epsfig}
\usepackage{multirow}
\expandafter\let\csname equation*\endcsname\relax
\expandafter\let\csname endequation*\endcsname\relax
\usepackage{amsmath}
\usepackage{relsize}
\usepackage{cite}

\usepackage{iopams}
\usepackage{color}  

\begin{document}

\title[Radiative nucleon capture with quasi-separable potentials]{Radiative nucleon capture with quasi-separable potentials}

\author{Shubhchintak$^1$, C A Bertulani$^{1,2}$, A M Mukhamedzhanov$^3$ and A T Kruppa $^4$}

\address{$^1$ Department of Physics, Texas A\&M University-Commerce, Commerce, TX 75429, USA}
\address{$^2$ Department of Physics and Astronomy, Texas A\&M University, College Station, Texas 77843, USA}
\address{$^3$ Cyclotron Institute, Texas A\&M University, College Station, TX 77843, USA}
\address{$^4$ Institute for Nuclear Research, Hungarian Academy of Sciences, Debrecen, PO Box 51, H-4001, Hungary}

\ead{\mailto{shub.shubhchintak@tamuc.edu}, \mailto{carlos.bertulani@tamuc.edu}, \mailto{akram@comp.tamu.edu},\mailto{atk@atomki.mta.hu}}

\begin{abstract}
We study radiative capture reactions using quasi-separable potentials. This procedure allows an easier treatment of non-local effects that can be extended to three-body problems. Using this technique, we calculate the neutron and proton radiative capture cross sections on $^{12}$C. The results obtained are shown to be in good agreement with the available experimental data. 
\end{abstract}

\maketitle
\section{Introduction}
\label{intro}
Separable potentials have been widely used in the literature to study nucleon-nucleon as well as nucleon-nucleus scattering problems \cite{yamaguchi,phillips,thomas,haiden,cattapan1,cattapan2,miyagawa}. Short range nuclear interactions by nature are non-local and can be treated in separable forms. However, this non-locality is often weak and therefore equivalent local potentials, e.g., of Woods-Saxon type, can be used to replace non-local potentials to a good extent \cite{pisent74}.  
Although separable potentials have found extensive use  in nuclear physics, they have also been used in other branches of physics. In nuclear physics their main application is in the realm of three or few-body scattering problems \cite{alt_02, alt_07}, where local potentials increase computational efforts. Actually, in three body problems one has to find the ``off-energy-shell" scattering amplitude, which can be handled more easily and in a more transparent manner, with separable potentials. Separable potentials help to convert the three-body  into effective two-body problems \cite{alt_07}, thus simplifying the problem. 

Our interest in this work is to invoke separable potentials to calculate cross sections for direct radiative nucleon capture reactions of importance for nuclear astrophysics. We make use of experimental phase shifts obtained from nucleon-nucleus scattering considering non-local potentials of separable form. We will closely follow the quasi-separable potential technique developed in Refs. \cite{cattapan1, cattapan2}. In the literature there is a large number of separable potentials available which have been proven to work very well in different energy ranges. We refer to at least three of these given in Refs, \cite{cattapan1, cattapan2,miyagawa,alt_07}. The first potential set given in Refs. \cite{cattapan1, cattapan2} explains very well nucleon+$^{12}$C experimental phase shifts for $L\leq2$  in the low energy range (below $\sim 5-7$ MeV), whereas the second potential set of Ref. \cite{miyagawa} yields good agreement with experiments at higher energies. The third set, used in Ref. \cite{alt_07}, is constructed in such a way so as to improve shortcomings observed in the first two sets to a satisfactory extent. 

As we are interested in astrophysical reactions, which take place at low energies, the first set of potential is suitable for our purposes. Furthermore, the complication in the case of charge-particle scattering can be taken care of in a simple and accurate manner using this potential. The potential parameters required for proton-nucleus scattering are almost the same as those for neutron-nucleus scattering. Therefore, as an application of the method we calculate direct neutron and proton capture cross sections for $^{12}$C(n,$\gamma$)$^{13}$C and $^{12}$C(p,$\gamma$)$^{13}$N reactions.

The nuclei under interest, $^{13}$C and $^{13}$N are very important in astrophysical scenarios. They are formed in the CNO cycle, where a proton capture on $^{12}$C (following the triple-$\alpha$ process) leads to $^{13}$N, which then decays by positron emission and forms $^{13}$C \cite{wiescher99, rolfs_book}. This is also known as the cold CNO cycle and takes place in massive main-sequence stars in the reaction chain $^{12}$C(p,$\gamma$)$^{13}$N($\beta^+, \nu$)$^{13}$C(p,$\gamma$)$^{14}$N(p,$\gamma$)$^{15}$O($\beta^+, \nu$)$^{15}$N(p$,\alpha$) $^{12}$C. However, in high temperature conditions occurring in novae and x-ray bursts, the $^{13}$N formed after the first step undergoes proton capture and yields a new cycle, known as the hot CNO cycle $^{12}$C(p$,\gamma$)$^{13}$N(p$,\gamma$)$^{14}$O($\beta^+, \nu$)$^{14}$N(p$,\gamma$)$^{15}$O($\beta^+, \nu$)$^{15}$N(p$,\alpha$) $^{12}$C. $^{13}$C also leads to neutron formation and acts as a neutron source in helium burning asymptotic giant branch stars by  means of the $^{13}$C($\alpha,$n)$^{16}$O reaction \cite{straniero}. In neutron rich environments, $^{12}$C can also undergo neutron capture processes and  start a neutron induced CNO cycle that leads to the formation of other heavier elements and hence a breakout of the CNO cycle. 

Neutron capture cross sections on $^{12}$C have been measured at thermal as well as at neutron energies between 10 and 550 MeV \cite{mugha_82,nagai_91,ohsaki,kikuchi}. The cross section has contributions from four low-lying states of $^{13}$C and deviates from the usual $1/v$ law \cite{ohsaki}. Its energy dependence has been studied within different theoretical models \cite{ho, mengoni, lin, baye} and it has been found that the capture is dominated by the transition to the first excited state of $^{13}$C, which is a weakly-bound halo state \cite{otsuka}. The $^{12}$C(p,$\gamma$)$^{13}$N cross sections have also been measured at long energy interval $70-2500$ keV \cite{lamb, young, rolf, burtebaev} and are well explained theoretically using direct capture plus resonant contributions. In fact, many different direct capture models have been used to calculate the nucleon capture cross sections which differ in the way one chooses initial and final states. In this context, the present work adds an additional method to obtain the nucleon capture reactions. Such reactions are of enormous interest for nuclear astrophysics (see, e.g. Ref. \cite{BK16}). As separable potentials are easier to handle non-local effects, they would enable improvements in numerical calculations of cross sections used to determine element production in several astrophysical scenarios involving both two-body and three-body reactions. Another goal of this work is to test the behavior of the continuum wavefunction obtained with the adopted separable potential so as to use it in Faddeev-AGS equations for three-body systems.

This paper is organized in the following way. In section 2, we briefly describe our formalism to use non-local nucleon-nucleus potentials to obtain a feasible calculation of nucleon capture cross sections. In section 3, we present our results, including a phase shift analysis to reproduce experimental results. Our conclusions are presented in section 4. 

\section{Formalism}
\subsection{Radiative capture in the potential model}
\label{sec:2}
For the direct radiative capture reaction $b + c \rightarrow a + \gamma$, where $c$ is a nucleon, the cross section is given by (see, e.g., Refs. \cite{Ber03,huang,MSB16})
\begin{eqnarray}
\sigma^{cap}_{\pi L}&=&\frac{(2\pi)^3}{k^2}\left(\frac{E_\gamma}{\hbar c}\right)^{2L+1}\frac{2(2I_a + 1)}{(2I_b+1)(2I_c+1)}\frac{L+1}{L[(2L+1)!!]^2}\nonumber\\
&\times& \sum_{l,j,J}(2J+1)\left\{\begin{array}{ccc} j & J & I_b\\ J_0 & j_0 & L\end{array}\right\}^2 |\langle l j||O_{\pi L}||l_0 j_0 \rangle|^2,   \label{a1}
\end{eqnarray}
where, $\pi L$ stands for electric or magnetic transition of multipolarity $L$, and $E_\gamma = E_{c.m.} + \epsilon$, with $E_{c.m.}$ being the center of mass energy of the $b-c$ system with binding energy $\epsilon$. $k$ is the momentum corresponding to $E_{c.m.}$ and $I_a$, $I_b$ and $I_c$ are the intrinsic spins of the respective particles. 

Here, the relative angular momentum $l$ of the pair $b-c$ couples with the spin of nucleon $I_c$ to yield $j$, so that ${\bf l} + {\bf I_c} = {\bf j}$. This then couples with the intrinsic spin of the ``core" $I_b$ to yield the initial total angular momentum  ${\bf J} ={\bf j} + {\bf I_b}$. $l_0$, $j_0$ and $J_0$ denote the corresponding bound state quantum numbers. In Eq. \eqref{a1}, $\langle l j||O_{\pi L}||l_0 j_0 \rangle$ is the reduced matrix element. For electric multipole transitions, which is the case of our interest, it is given by \cite{Ber03,huang} (see also Eq. A2.23 of Ref. \cite{Law80})
\begin{eqnarray}
\langle l j||O_{\pi L}||l_0 j_0 \rangle &=&  \delta_{l+l_0+L,even}(-1)^{(l_0+l-j+L-1/2)} \frac{e_L}{\sqrt{4\pi}} \sqrt{(2L+1)(2j_0+1)}\nonumber\\
&\times&\left(\begin{array}{ccc}j_0 & L & j\\1/2 & 0 & -1/2\end{array}\right) \int_0^\infty dr r^L u_{l_0j_0}^{J_0}(r)u_{lj}^{J}(r), \label{a2}
\end{eqnarray}
where $e_L$ is the effective charge given by $e_L = Z_b e(-m_c/m_a)^L + Z_c e(m_b/m_a)^L$, with $m_i$ and $Z_i$ being the masses and charges of the respective particles. $u_{l_0j_0}^{J_0}(r)$ and $u_{lj}^{J}(r)$ are the bound state and continuum wave function, respectively.

Here, there is no advantage of using a separable potential and the bound state wave function can be calculated by solving the Schr{\"o}dinger equation with a Woods-Saxon (WS) potential, where the potential depth is adjusted to get the corresponding binding energy of the state. We have verified this assertion by using the result of the potential model as described above, and compared to calculations for the bound state using the separable potential defined below using the code PSEUDO \cite{KP85}. As shown, e.g., in Ref. \cite{huang}, it is necessary to change the WS potential parameters to calculate the continuum wavefunctions and to reproduce in a reasonable fashion the experimental values of the radiative cross sections and astrophysical S-factors.  Instead, here we calculate the continuum wave function to obtain the proper scattering matrix (S-matrix) for nucleon-nucleus scattering using the quasi-separable potential technique described in Ref. \cite{cattapan1, cattapan2}. The advantage of the method is that it allows us a consistent reproduction of the experimental phase shifts and at the same time  the radiative capture cross sections, using separable potentials for both situations. Moreover, separable potentials allow an easy implementation of three-body calculations which is our next goal.  We will briefly discuss how to implement S-matrix calculations with this formalism for  neutron-nucleus and  proton-nucleus scattering.

\subsection{S-matrices and wavefunctions from quasi-separable potentials}

If we represent the initial state of the $b+c$ system in partial wave form as $|\alpha l jJM\rangle$, where $\alpha$ is the channel number representing the core state so that $I_{\alpha} = I_b$ (with the convention that $\alpha = 1$ represents the elastic channel) and $l, j, J$ are the quantum numbers as explained above with $M$ being the z-component of total angular momentum $J$, then the nuclear interaction in a quasi-separable form can be written as 
\begin{eqnarray}
V = -\sum_{\substack{\alpha l j,\beta l' j'\\JM}}|{\bf \chi};\alpha l j JM\rangle \lambda_{\alpha l j,\beta l' j'}^{J\Pi}\langle\beta l' j' JM;{\bf \chi}|, \label{a3}
\end{eqnarray}
where $l', j', J'$ and $M'$ are quantum numbers of the system in channel $\beta$. $\lambda$ represents the coupling between the channels and the coupling matrix is considered to be symmetric i.e., $\lambda_{\alpha l j,\beta l' j'}^{J\Pi}$ = $\lambda_{\beta l' j', \alpha l j}^{J\Pi}$. This assures that the potential is Hermitian and separability has been considered only in the relative motion variable $r$ in each channel. $|{\bf \chi};\alpha l j JM\rangle$ denotes form factor vectors and in momentum space its representation is
\begin{eqnarray}
\langle\alpha l j JM;{k}|{\bf \chi}; \beta l' j' J'M'\rangle \nonumber &=&  \delta_{\alpha \beta}\delta_{l l'}\delta_{jj'}\delta_{JJ'}\delta_{MM'} i^{-l}u_{\alpha l j}^J(k),
~~~~ \label{a4}
\end{eqnarray}
where the form factor $u_{\alpha l j}^J(k)$, is given by
\begin{eqnarray}
u_{\alpha l j}^J(k) &=& \sqrt{\frac{2}{\pi}} (2l)!!g_{\alpha l j}^J(k),   \hspace{0.3cm}  {\rm with} \ \ \ \ \  g_{\alpha l j}^J(k)=\frac{k^l}{[k^2+(\mbox{\boldmath$\beta$}_{\alpha l j}^J)^2]^{(l+1)}}.  
~~~ \label{a5}
\end{eqnarray}
$\mbox{\boldmath$\beta$}_{\alpha l j}^J$ is the fitting parameter having units of inverse length, with the assumption $\mbox{\boldmath$\beta$}_{\alpha l j}^J$ = $\mbox{\boldmath$\beta$}_{lj}$. 

One can write the S-matrix as
\begin{eqnarray}
{S}_{\alpha l j,\beta l' j'}^{J\Pi} &=& \delta_{\alpha\beta}\delta_{ll'}\delta_{jj'} + i^{l'-l+1}\pi \mu k_{\alpha}^{1/2}u_{\alpha l j}^J(k_{\alpha}){\cal T}_{\alpha l j,\beta l' j'}^{J\Pi}u_{\beta l' j'}^J(k_{\beta})k_{\beta}^{1/2},\label{a6}
\end{eqnarray}
where ${\cal T}_{\alpha l j,\beta l' j'}^{J\Pi}$ is a matrix related to T-matrix as
\begin{eqnarray}
{T}_{\alpha l j,\beta l' j'}^{J\Pi}(k,k') = -i^{l'-l}~u_{\alpha l j}^J(k) {\cal T}_{\alpha l j,\beta l' j'}^{J\Pi} u_{\beta l' j'}^J(k'), \label{a7}
\end{eqnarray}
satisfying the following coupled equation,
\begin{eqnarray}
{\cal T}_{\alpha l j,\beta l' j'}^{J\Pi} = \lambda_{\alpha l j,\beta l' j'}^{J\Pi} - \sum_{\gamma l'' j''} \lambda_{\alpha l j,\gamma l'' j''}^{J\Pi} {\cal G}_{\gamma l'' j''}^{J\Pi}{\cal T}_{\gamma l'' j'',\beta l' j'}^{J\Pi}, \label{a8}
\end{eqnarray}
with
\begin{eqnarray}
{\cal G}_{\gamma l'' j''}^{J\Pi} &=& \langle\gamma l'' j'' JM;{\bf \chi}|G_0|{\bf \chi};\gamma l'' j'' JM\rangle 
=\mu\int_0^{\infty}\frac{[u_{\gamma l''j''}(k)]^2 k^2}{k_{\gamma}^2-k^2+i\epsilon}dk. \label{a9}
\end{eqnarray}

In the above equation $G_0 (z)=(z-H_0)^{-1}$ is the resolvent for unperturbed Hamiltonian $H_0$. Eq. (\ref{a6}) is the $S$-matrix in a quasi-separable potential form for the case when $c$ is a neutron and there is no Coulomb interaction. In the case of proton-nucleus scattering, one has to take care of the Coulomb potential ${V_c}$ along with the nuclear potential $V$. The transition operator can be written as the sum of the pure Coulomb ($t$) and Coulomb modified  nuclear ($\tau$) transition operators, i.e $ T = t + \tau$, where $\tau$ is given by
\begin{eqnarray}
\tau=(1+{V_c}G_c)T_c(1+G_cV_c), 
\end{eqnarray} 
with $G_c(z) = (z-H_0-{V_c})^{-1}$ being the Coulomb-perturbed resolvent and $T_c$ satisfies the equation $T_c = V +VG_cT_c$.

The reduced $S$-matrix expression in this case has a similar form as in Eq. (\ref{a6}), with all the quantities replaced by their Coulomb corrected form.  That is,
\begin{equation}
{S}_{C,\alpha l j,\beta l' j'}^{J\Pi} = \delta_{\alpha\beta}\delta_{ll'}\delta_{jj'} + i^{l'-l+1} 
 \pi \mu k_{\alpha}^{1/2}u_{C,\alpha l j}^J(k_{\alpha}){\cal T}_{C,\alpha l j,\beta l' j'}^{J\Pi}u_{C,\beta l' j'}^J(k_{\beta})k_{\beta}^{1/2}, \label{a10}
\end{equation}
where the Coulomb corrected matrices, ${\cal T}_C$ and ${T}_C$ satisfy a similar relation as in Eq. (\ref{a7}), i.e.,
\begin{eqnarray}
{T}_{C,\alpha l j,\beta l' j'}^{J\Pi}(k,k') = -i^{l'-l}~e^{i\sigma_l(k)}~u_{C,\alpha l j}^J(k) 
{\cal T}_{C,\alpha l j,\beta l' j'}^{J\Pi} u_{C,\beta l' j'}^J(k') e^{i\sigma_{l'}(k')}. \label{tct}
\end{eqnarray}

In the above equation $\sigma_l$ is the Coulomb phase shift and $u_{C,\alpha l j}^J(k)$ is the Coulomb corrected nuclear form factor given by,
\begin{eqnarray}
u_{C,\alpha l j}^J(k)&=&\sqrt{\frac{2}{\pi}} (2l+1)!!C_l(\eta) g_{\alpha l j}^J(k) \exp{\left[2\eta {~} \arctan\left(\frac{k}{\mbox{\boldmath$\beta$}_{\alpha l j}^J}\right)\right]},\label{a11}
\end{eqnarray}
where $\eta$ is the Coulomb parameter and the functions $C_l(\eta)$ are given by,
\begin{eqnarray}
C_l(\eta)=\frac{{\sqrt{l^2+\eta^2}}}{l(2l+1)}~C_{l-1} (\eta);~~~~~~ C_0(\eta)={\sqrt\frac{2\pi\eta}{e^{2\pi\eta}-1}}  
\end{eqnarray}
For more detailed derivations of the quantities described above one is referred to Refs. \cite{cattapan1,cattapan2}.

The elastic component of the $S$-matrix, i.e., ${S}_{1 l j,1 l' j'}^{J\Pi}$, is used to obtain the phase shifts $\delta_{lJ}$ which are related by $S=e^{2i\delta}$. Two steps are needed; the first one is to find the possible values of the parameters $\mbox{\boldmath$\beta$}$ and $\lambda$  that reproduce all the resonances at right positions. These values are further improved by fitting to the experimental phase shifts. 
Once the potential parameters are all set, we solve the Schr{\"o}dinger equation to obtain the scattering wave function which asymptotically behaves as,
\begin{eqnarray}
u_{lj}^{J}(r)(r\rightarrow \infty) = -\sqrt{\frac{2}{\pi}} ~ e^{i\delta_{lJ}}\left[F_l \cos(\delta_{lJ})+G_l \sin(\delta_{lJ})\right],\label{a12}
\end{eqnarray}
where $F_l$ and $G_l$ are regular  and irregular Coulomb wave functions.

\section{Results and discussions}
\subsection{Phase-shifts}
Neutron and proton radiative capture on $^{12}$C lead to the formation of $^{13}$C and $^{13}$N, respectively. Given that the total neutron capture cross section has contributions from four low-lying states of $^{13}$C \cite{ohsaki}, we calculate the neutron capture cross section to the states with spin parity $J^{\pi}=1/2^-$ for the ground state, $1/2^+$ for the $1^{st}$ excited state (e.s.), $3/2^-$  for the $2^{nd}$ e.s., and $5/2^+$ for the $3^{rd}$ e.s., respectively. These states are formed by coupling of the $0^+$ ground state of $^{12}$C with the neutron in the 1$p_{1/2}$, 2$s_{1/2}$, 1$p_{3/2}$ and 1$d_{5/2}$ orbital with one neutron removal energy, $S_n$, equal to 4.95, 1.86, 1.27 and 1.09 MeV, respectively. The proton capture cross section on the other hand is only calculated for capture to the ground state of $^{13}$N with $J^{\pi}= 1/2^-$, formed by coupling the 1$p_{1/2}$ proton to the $0^+$ ground state of $^{12}$C, with a binding energy of $S_p= 1.94$ MeV. In order to calculate the bound state wave functions of $^{13}$C, we use the Woods Saxon parameters of Ref. \cite{mengoni},  i.e., $r_0 = 1.236$ fm, $a = 0.62$ fm and $V_{so} = -7$ MeV. The values of the potential depth required to reproduce the binding energy for the states $1/2^-$, $1/2^+$, $3/2^-$ and $5/2^+$ are 41.35, 56.90, 28.81 and 56.85 MeV, respectively. For $^{13}$N, we use $r_0 = 1.25$ fm, $a = 0.65$ fm and $V_{so} = -10$ MeV (same as in Ref. \cite{huang}) and the potential depth in this case is 41.65 MeV. We adopted the WS potentials here for convenience, to implement our radiative capture numerical calculations using the code RADCAP \cite{Ber03}. As we have stated before, for bound states, equivalent separable potentials can be found leading basically to the same wavefunctions for these reactions. We thus emphasize that our calculations are consistent with the use of separable potentials for bound and continuum states. 

\begin{table}[ht]
\begin{center}
\caption{Fitting parameter $\mbox{\boldmath$\beta$}_{lj}$. They are the same for both $^{13}$C and $^{13}$N.   }
\begin{tabular}{|c|c|c|}
\hline\hline

$l$&$j$  &  $\mbox{\boldmath$\beta$}_{lj}$ (fm$^{-1}$)    \\ 
\hline
0& {1}/{2} & 0.57 \\
1& {1}/{2} & 1.76 \\
1& {3}/{2} & 1.22 \\
2& {3}/{2} & 2.16 \\
2& {5}/{2} & 2.96 \\
\hline
\hline
\end{tabular}
\end{center}
\end{table}

\begin{table}[ht]
\begin{center}
\caption{Strength parameter $\lambda_{\alpha l j, \beta l' j'}^{J\pi}$ used in Eq. \ref{a3}, for $^{13}$C and $^{13}$N. }
\begin{tabular}{|c|c|c|c|}
\hline\hline
Phase shift  & $\alpha ~l ~j; \beta ~l'~j'$   &  {$^{13}$C} & {$^{13}$N}  \\ 
\hline
          &      &  $\lambda_{\alpha l j, \beta l' j'}^{J\pi}$ & $\lambda_{\alpha l j, \beta l' j'}^{J\pi}$  \\
\hline 
$s_{1/2}$ &  $1 0 \frac{1}{2}; 1 0 \frac{1}{2}$ & 0.8464  & 0.9801\\
\hline
$p_{1/2}$ &  $1 1 \frac{1}{2}; 1 1 \frac{1}{2}$ & 73.445  &       \\
\hline
$p_{3/2}$ &  $1 1 \frac{3}{2}; 1 1 \frac{3}{2}$ & 98.8    &        \\
\hline
$d_{3/2}$ &  $1 2 \frac{3}{2}; 1 2 \frac{3}{2}$ & 268.9   &  277.2 \\
\hline
$d_{5/2}$ &  $1 2 \frac{5}{2}; 1 2 \frac{5}{2}$ & 2684    &  2770 \\
\hline
\hline
\end{tabular}
\end{center}
\end{table}

For both of these capture reactions the E1 multipolarity dominates \cite{rolf,ohsaki,huang}. Hence, we use proper initial angular momenta states in the continuum leading to E1 transitions to the corresponding bound states of each nucleus. For transitions to the ground state of $^{13}$C, the  proper continuum states have $s_{1/2}$ and $d_{3/2}$ angular momenta, for the  $1^{st}$ e.s. they are $p_{1/2}$ and $p_{3/2}$, for the $2^{nd}$ e.s. they are $s_{1/2}$, $d_{3/2}$ and $d_{5/2}$ and for the $3^{rd}$ e.s. it is just the $p_{3/2}$ angular momentum, respectively. Similarly, for E1 transitions to the ground state of $^{13}$N, the continuum states taken to be $s_{1/2}$ and $d_{3/2}$. 

The calculations of continuum wave functions depend upon the phase shifts and hence the scattering matrix, which are obtained by the procedure given in the previous section.  As we need only elastic component of S-matrix to calculate the phase shifts, we will only use one single channel, i.e., only the ground state of the target ($^{12}$C), to calculate the S-matrix.  A multi-channel formalism could be taken in order to study the effect of inclusion of more target states on the capture cross sections, provided the fit parameters would be known for all channels under consideration. 
We use the same set of parameters $\mbox{\boldmath$\beta$}$, $\lambda$ as those obtained in Ref. \cite{cattapan1, cattapan2}. For two cases, discussed later in this section, we employ changes in order to get the best fit with the experimental data. We also generated our own best fit parameters for the $p_{3/2}$ state of $^{13}$C by fitting it to the experimental phase shift. In Tables I and II, we give the parameters $\mbox{\boldmath$\beta$}$ and $\lambda$, respectively, used to obtain the different phase shifts. 

\begin{figure}[h]
\centering
\includegraphics[trim=0.5mm 0.5mm 0.5mm 0.5mm,clip,width=7cm]{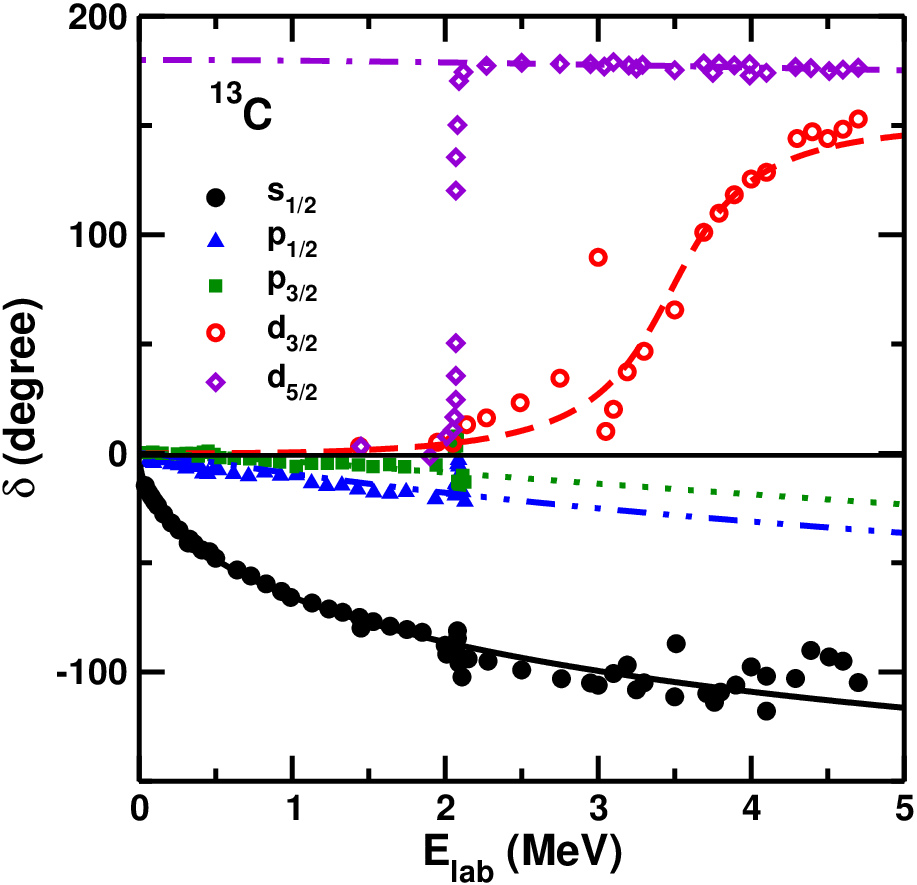}
\caption{\label{fig:1} $s, p, d$ phase shifts for n-$^{12}$C scattering. Solid, dotted-dotted-dashed, dotted, dashed and dotted-dashed lines represent the calculated $s_{1/2}$, $p_{1/2}$, $p_{3/2}$, $d_{3/2}$ and $d_{5/2}$ phase shifts, respectively, whereas filled circles, upper triangles, square boxes, open circles and diamond boxes represent the corresponding experimental data which are taken from Refs. \cite{wills, lister, pisent}.}

\end{figure}

\begin{figure}[h]
\centering
\includegraphics[trim=0.5mm 0.5mm 0.5mm 0.5mm,clip,width=7cm]{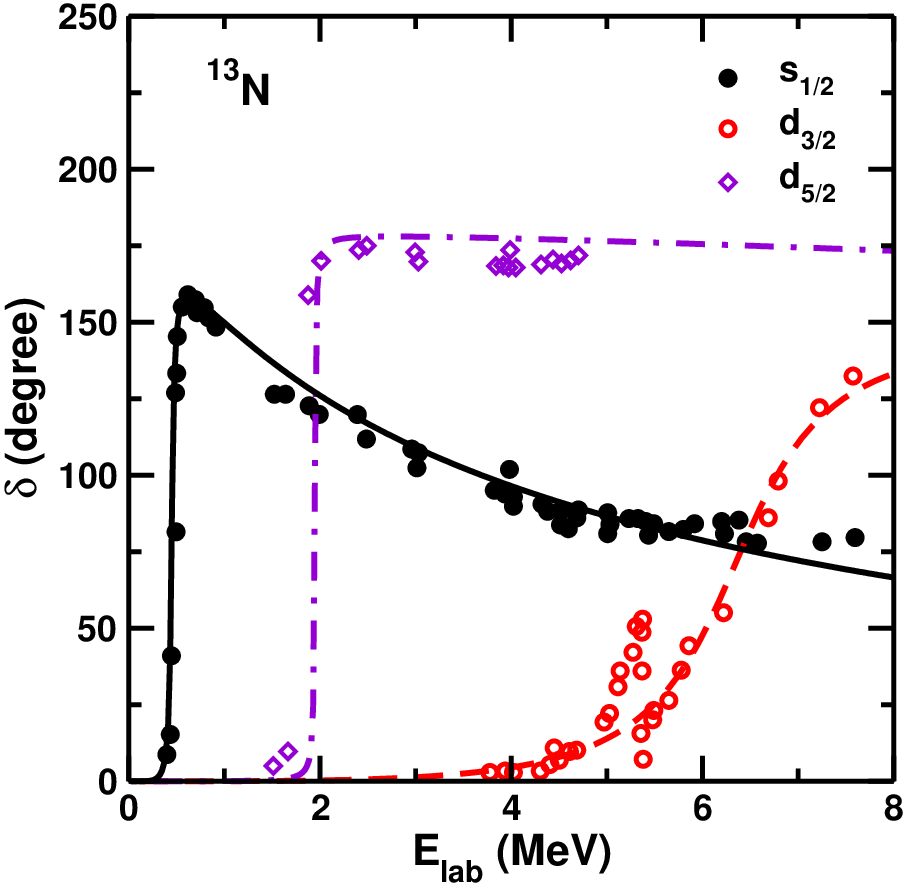}
\caption{\label{fig:2} $s, d$ phase shifts for p-$^{12}$C scattering (solid lines). The symbols and lines are the same as in Fig. \ref{fig:1}. The experimental data are taken from Refs. \cite{reich,moss,jackson,bernstein,trachslin,drigo}.}
\end{figure}

The results of our phase shift calculations for $^{13}$C and $^{13}$N are presented in Figs. \ref{fig:1} and \ref{fig:2}, respectively. In both of these figures we plot the phase shifts $\delta$, in degrees, as a function of the laboratory energy $E_{lab}$. 
In these figures solid, dotted-dotted-dashed, dotted, dashed and dotted-dashed lines represent the calculated $s_{1/2}$, $p_{1/2}$, $p_{3/2}$, $d_{3/2}$ and $d_{5/2}$ phase shifts, respectively, whereas filled circles, upper triangles, square boxes, open circles and diamond boxes represent the corresponding experimental data  taken from Refs. \cite{wills, lister, pisent} for n-$^{12}$C scattering and from Refs. \cite{reich,moss,jackson,bernstein,trachslin,drigo} for p-$^{12}$C scattering, respectively.
For both of these nuclei, we are able to reproduce the results presented in Ref. \cite{cattapan1}. The calculated phase shifts match well the experimental data. Furthermore, some narrow resonances for $d_{5/2}$ and $d_{3/2}$ states, both for proton and neutron scattering are found to be of compound nature and can only be explained by considering target excitations to the $2^+$ state as in Ref. \cite{cattapan2}. 
We also want to point out that for the proton case, the parameter $\mbox{\boldmath$\beta$}_{0 \frac{1}{2}}$ given in Table I is slightly corrected to 0.5723 fm$^{-1}$ to yield a 1/2$^{+}$ resonance at the right position, i.e., at 0.422 MeV in the c.m., and for a better fit of the experimental phase-shift data,  as seen in Fig. 2. A slight change is also needed for $\mbox{\boldmath$\beta$}_{1 \frac{1}{2}}$ in the case of $^{13}$C, where it is modified from 1.72 fm$^{-1}$ to 1.76 fm$^{-1}$ in order to obtain a best fit.

\subsection{Capture cross sections}
Once the phase shifts have been obtained, we now proceed to calculate the continuum wave functions with the asymptotic behavior described in Eq. (\ref{a12}).  We calculate the neutron and proton capture cross sections using Eqs. (\ref{a1}) and (\ref{a2}), as all the quantities are now established.

\begin{figure}[h]
\centering
\includegraphics[trim=0.5mm 0.5mm 0.5mm 0.5mm,clip,width=9cm]{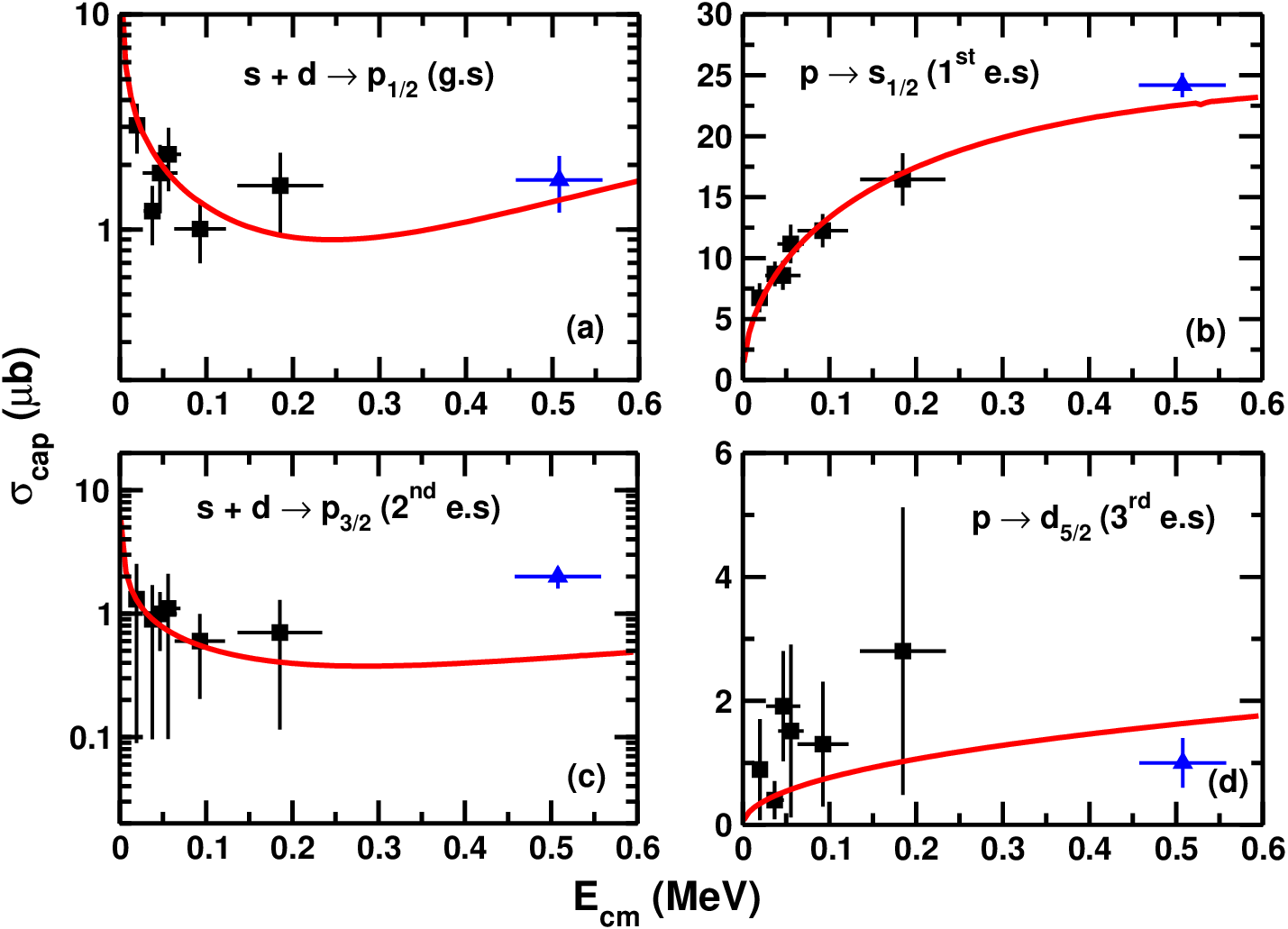} 
\caption{\label{fig:3} Radiative neutron capture cross sections for capture to the four low-lying states in $^{12}$C. Filled Square boxes and filled triangles are the experimental data from Refs. \cite{ohsaki} and \cite{kikuchi}, respectively. }
\end{figure}

\begin{figure}[h]
\centering
\includegraphics[trim=0.5mm 0.5mm 0.5mm 0.5mm,clip,width=7cm]{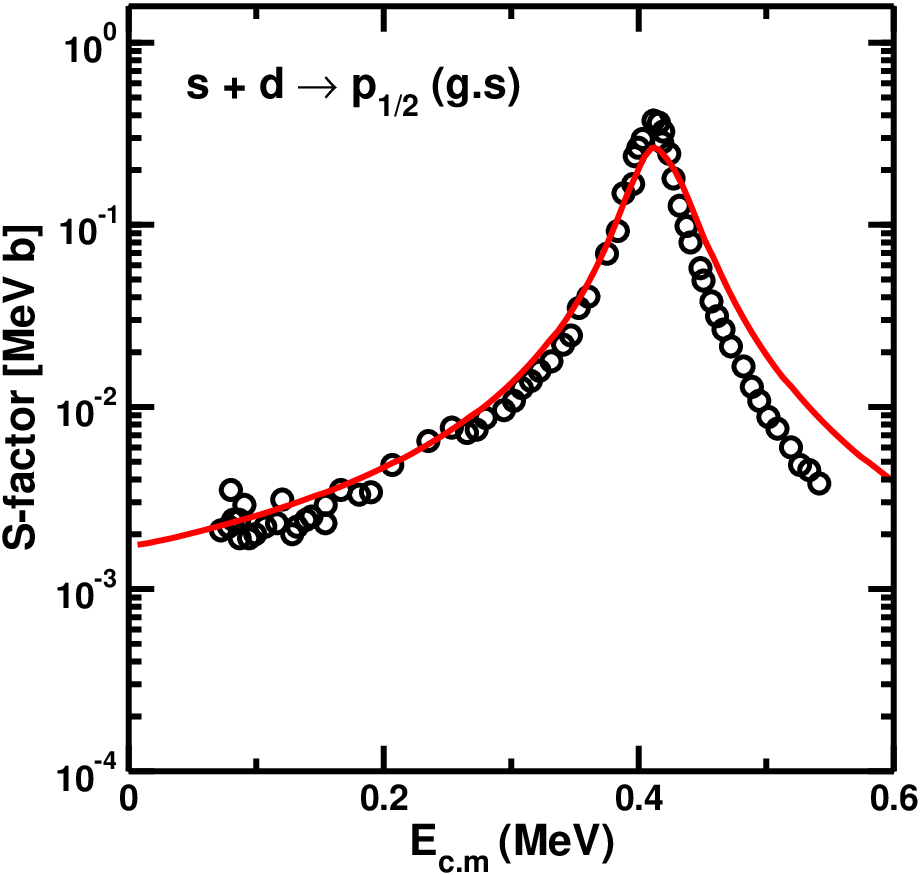}
\caption{\label{fig:4} Astrophysical S-factor for the $^{12}$C($p$, $\gamma$)$^{13}$N reaction. Experimental data are from Ref. \cite{rolf}. }
\end{figure}

Given that the continuum wave functions are obtained from phase shifts  that reproduce experimental data, we expect that our numerical calculations for capture cross sections are more reliable than standard potential models. Fig. \ref{fig:3} shows our calculations of neutron capture cross sections (solid line) for capture to different states of $^{13}$C compared with respective experimental data. The square boxes in all figures represent data from Ref. \cite{ohsaki}, whereas the upper triangles correspond to experimental data from Ref. \cite{kikuchi} where the cross sections were measured only at the 0.55 MeV laboratory energy. The spectroscopic factor ($C^2S$) used for the ground state, the $1^{st}$ excited state, the $2^{nd}$ excite state  and the $3^{rd}$ excited state are 0.77, 1.0, 0.14, 0.58, respectively. They have been taken from Ref. \cite {ohnuma}, except for the $1^{st}$ e.s. where $C^2S$ equal to 1.0 in our case. This value explains the data very well. It is clear from the figure that the cross section obtained by this method are in good agreement with the experimental data, especially at low energies.

Fig. \ref{fig:4} shows the S-factor for proton capture to the ground state of $^{13}$N (solid line). The experimental data (open circles) are from Ref. \cite{rolf}. The spectroscopic factor $C^2S$  in this case was chosen to be 0.21, which is quite different than the one used for ground state of  $^{13}$C. As mentioned earlier, the E1 transition to the ground state of $^{13}$N (1/2$^{-}$) takes place from $s$ or $d$ continuum wave. The dominant contributions to the cross sections come from the $s$ continuum wave, where there is a resonance at 0.422 MeV, as mentioned earlier. This resonance yields a peak in the cross section which can be seen in the Fig. \ref{fig:4}. Clearly, our calculations again yield a good agreement with the experimental data over the present energy range.

\section{Conclusions}
We have discussed the utility of quasi-separable potentials to describe nucleon-nucleus scattering and applications to nuclear astrophysics. We have shown how this method can be used to calculate radiative capture cross sections for astrophysical important reactions. As an example, we have considered the case of neutron and proton scattering from $^{12}$C, where the scattering phase shifts are used to infer  the scattering wave functions. The scattering wave functions corresponding to the scattering potential parameters determined from phase shifts analysis of the experimental data are used as input to calculate  neutron and proton capture cross sections. Our results are in good agreement with the available experimental data in the energy range considered in this work.

This technique can be used to obtain the continuum wave functions with a better description of phase-shifts and radiative capture cross sections than the usual potential models.  A limitation of using WS or separable potential models (or even any microscopic model) is that, to be consistent, one also needs to reproduce scattering phase shifts or elastic scattering cross sections. While in the WS and separable potential models this is an easy task, just amounting to redefine the potential parameters, in microscopic models it can be a formidable task. The advantage of using separable potentials over WS potential models is the ability to include non-local features. Non-local dependence is of relevance for the treatment of three-body systems, as well as simplifying the calculations enormously. The multi-channel formalism  briefly discussed in this paper, can also be used to study the effects of inclusion of numerous target states on the capture cross sections. The use of a single channel is limited to lower mass nuclei, up to $^{16}$O. With the increase of the nuclear mass, the nuclear structure becomes more complicated and one needs to take into consideration the effect of other channels in order to explain the resonances which also become of non-single-particle nature. 
Applications of this technique to deuteron-induced reactions are in progress.

\section{Acknowledgements}
C.A.B. Acknowledges support from the U.S. NSF Grant number 1415656 and the U.S. DOE Grant number DE-FG02-08ER41533. A.M.M.  acknowledges support from the U.S. DOE grant numbers  DE-FG02-93ER40773 and DE-FG52-09NA29467 and  by the U.S. NSF grant number PHY-1415656. A.T.K. acknowledges support from the Hungarian Scientific Research Fund-OTKA, grant No. K112962.

\end{document}